\documentclass[
english, aps, prd,
superscriptaddress,
preprintnumbers,
nofootinbib,
twocolumn
]{revtex4-2}

\usepackage[utf8]{inputenc}

\usepackage{amssymb}
\usepackage{amsmath}
\usepackage{amsfonts}
\usepackage{graphicx}
\usepackage{siunitx}
\DeclareSIUnit\barn{b}

\usepackage{babel}
\usepackage{setspace}

\usepackage[colorlinks,linktocpage=true]{hyperref}
\hypersetup{citecolor=blue,linkcolor=blue,urlcolor=blue}

\newcommand{\be}{\begin{equation}}
\newcommand{\ee}{\end{equation}}
\newcommand{\MSbar}{\ensuremath{\overline{\text{MS}}}}
\newcommand{\qtyrangemath}[3]{\qtyrange[range-units=single,range-phrase={-}]{#1}{#2}{#3}}

\newcommand{\commonPlotAnnotation}{The LO and NLO distributions are rescaled by the $K$-factor $K_i=\sigma_{\mathrm{NNLO}{\times}(\mathrm{MiNNLO{+}PS}})/\sigma_i$, so that all the shown distributions have the same integral. The shown predictions correspond to the central choices of the renormalization and factorization scales, which are described in the main text. The bottom panels show the relative deviation from the rescaled $\mathrm{LO}{\times}(\mathrm{LO{+}PS})$ distribution. The bands show the Monte-Carlo uncertainty only.}

\allowdisplaybreaks

\begin{document}

\preprint{TTP25-022, P3H-25-047, MPP-2025-126}

\title{Parton-shower and fixed-order QCD effects in Higgs-boson production in weak-boson fusion and its decays to bottom quarks}

\def\KIT{Institute for Theoretical Particle Physics,
	Karlsruhe Institute of Technology, 76128 Karlsruhe, Germany}
\def\MPP{Max-Planck-Institut für Physik,
    Boltzmannstraße 8, 85748 Garching, Germany}
\def\TUM{Physik Department T31, Technische Universität München,
    James-Frank-Straße 1, 85748 Garching, Germany}

\author{Arnd~Behring}
\email{abehring@mpp.mpg.de}
\affiliation{\MPP}

\author{Kirill~Melnikov}
\email{kirill.melnikov@kit.edu}
\affiliation{\KIT}

\author{Ivan~Novikov}
\email{ivan.novikov@kit.edu}
\affiliation{\KIT}

\author{Giulia Zanderighi}
\email{zanderighi@mpp.mpg.de}
\affiliation{\MPP}
\affiliation{\TUM}

\begin{abstract}
\noindent
Recently, it was observed~\cite{Asteriadis:2024nbg} that an aggressive cut on the $b$-jets' transverse momenta applied to Higgs-boson production in weak-boson fusion followed by the decay $H \to b \bar b$, leads to very large QCD corrections to the fiducial cross section.
In this paper we show that these corrections are caused by soft and collinear QCD radiation and, therefore, can be efficiently treated by a parton shower. We combine the parton-shower description of the decay $H \to b \bar b$ with NNLO QCD corrections to Higgs production in weak-boson fusion and its subsequent decay, and demonstrate that the quality of the theoretical prediction is markedly improved even if $b$-jets with rather high transverse momenta are selected.
The remaining uncertainty of the theoretical prediction, mainly driven by imprecise modelling of $H \to b \bar b$ decay, is estimated to be of the order of $\mathcal{O}(\qtyrangemath{5}{7}{\percent})$.
\end{abstract}

\maketitle

%%%%%%%%%%%%%%%%%%%%%%%%%%%%%%%%%%%%%%%%%%%%%%%%%%

\section{Introduction}
\label{sec:introduction}

Studies of Higgs-boson production in weak-boson fusion (WBF) are an integral part of an effort to profile the Higgs boson with the highest precision, including measuring its couplings to electroweak gauge bosons and constraining its quantum numbers \cite{ATLAS:2022vkf,CMS:2022dwd}.
To maximize the number of available data samples, one considers Higgs-boson decay to $b$-jets, and uses characteristic kinematic features of the weak-boson fusion process to suppress QCD backgrounds.

However, in order to do that, one has to impose fairly aggressive cuts on the transverse momenta of the two $b$-jets \cite{ATLAS:2020bhl}. The analysis of Ref.~\cite{Asteriadis:2024nbg},
where such a fiducial cross section was computed through next-to-next-to-leading order (NNLO) in perturbative QCD, revealed very sizeable, $\mathcal{O}(\qty{-40}{\percent})$, perturbative corrections. In fact, the magnitude of QCD corrections was so large, and the convergence of the perturbative expansion so poor, that it was not clear how to assign a meaningful uncertainty to the computed fiducial cross section.

As discussed in Ref.~\cite{Asteriadis:2024nbg}, one can identify multiple independent sources of perturbative corrections, which all tend to reduce the fiducial cross section.
However, a particularly strong effect is related to the tendency of $b$-jets to lose transverse momentum to QCD radiation, which reduces the probability that a $b$-jet passes the high transverse-momentum selection cut. Since the value of the $p_\perp$-cut is high, $\mathcal{O}(m_H/2)$, very little energy needs to be radiated by a $b$-jet to lose enough transverse momentum to fail the cut, leading to a dramatic change in the cross section of this process at next-to-leading order, when QCD radiation off $b$-jets is encountered for the first time.
This observation led to a conjecture \cite{Asteriadis:2024nbg} that these large perturbative effects can be understood and properly described by promoting a parton-shower-aided description of $H \to b \bar b$ decay to the default ``leading-order'' prediction, and that higher-order QCD corrections should be computed on top of it.
We note that, since the Higgs boson is a color-neutral scalar particle, and since Ref.~\cite{Asteriadis:2024nbg} identifies the Higgs decay and the high value of the $b$-jet transverse momentum cut as the main source of large QCD effects, we may apply the parton shower to Higgs decay, but still use the standard perturbative description of the production process, where the QCD corrections are known to be modest \cite{Cacciari:2015jma,Cruz-Martinez:2018rod}. In this sense, we use the parton shower to resum large QCD effects caused by a high $p_\perp$-cut, rather than to generate unweighted events in a complicated process.

The approach described above requires the NNLO QCD corrections to Higgs-boson production in weak-boson fusion, for which we use the calculation of Ref.~\cite{Asteriadis:2021gpd}.
We also require the NNLO-accurate description of Higgs decay to $b$-jets matched to a parton shower. The corresponding generator was developed in Ref.~\cite{Bizon:2019tfo}.
It uses \texttt{Pythia8.3}~\cite{Bierlich:2022pfr}, the \texttt{POWHEG} framework~\cite{Nason:2004rx,Frixione:2007vw,Alioli:2010xd} and the \texttt{MiNLO} method~\cite{Hamilton:2012np,Hamilton:2012rf} to reweight generated Higgs-boson decay events to NNLO QCD accuracy.

The rest of the paper is organized as follows.
In Section~\ref{sec:details} we explain our setup in detail.
In Section~\ref{sec:results} we present the results of the calculation. We conclude in Section~\ref{sec:conclusions}.
%%%%%%%%%%%%%%%%%%%%%%%%%%%%%%%%%%%%%%%%%%%%%%%%%%

\section{Technical Details}
\label{sec:details}
We work in the narrow-width approximation, and factorize the Higgs-boson production in weak-boson fusion, and its subsequent decay into a $b \bar b$ pair, $H\to b\bar{b}$.
The differential cross section $\mathrm{d}\sigma$ of the combined process is written as
\be
    \mathrm{d}\sigma=\mathrm{d}\sigma_\textrm{WBF} \;\mathrm{Br}_{H\to b\bar{b}}
    \; \mathrm{d}\gamma_b\,,
    \label{eq1}
\ee
where $\mathrm{Br}_{H\to b\bar{b}}$ is the $H\to b\bar{b}$ branching ratio
\be
	\mathrm{Br}_{H\to b\bar{b}}=\frac{\Gamma_{H\to b\bar{b}}}{\Gamma_{H, \text{tot} }}\,,
\ee
$\mathrm{d}\sigma_\text{WBF}$ is the differential WBF production cross section,
and $\mathrm{d}\gamma_b$ is the ratio of differential and total $H\to b\bar{b}$ decay widths
\begin{equation}
	\mathrm{d}\gamma_b=\frac{\mathrm{d}\Gamma_{H\to b\bar{b}}}{\Gamma_{H\to b\bar{b}}}\,.
	\label{eq:gammab}
\end{equation}
We set the $H\to b\bar{b}$ branching ratio to a fixed value $\mathrm{Br}_{H\to b\bar{b}}=0.5824$~\cite{LHCHiggsCrossSectionWorkingGroup:2016ypw},
and consider different approximations for the differential production cross section $\mathrm{d}\sigma_\text{WBF}$ and the normalized differential decay rate $\mathrm{d}\gamma_b$.

The motivation for writing the cross section for the full process $pp \to H(b \bar b) + 2j$ as in Eq.~\eqref{eq1}, is that upon integration over the entire decay phase space, the cross section for the combined process reduces to the product of the WBF cross section and the $H \to b \bar b$ branching ratio, regardless of the approximation used to model the decay subprocess. Hence, the following equation holds
\be
	\int \limits_{\mathrm{decay}} \mathrm{d}\sigma=
	\mathrm{d}\sigma_\text{WBF}\, \mathrm{Br}_{H\to b\bar{b}}\,.
\ee
Furthermore, because the numerator and the denominator in Eq.~\eqref{eq:gammab} are always calculated in the same approximation, $\gamma_b$ does not depend on the overall normalization of the $H\to b\bar{b}$ decay rate.
As a result, the impact of different approximations on the cross section of the combined WBF process $pp \to H(b \bar b) + 2j$ is exclusively due to changes in the kinematics of decay events, and the ensuing changes in their acceptance rate.

In Ref.~\cite{Asteriadis:2024nbg} both the production and decay subprocesses were treated at fixed order, and higher-order terms arising in the product $\mathrm{d}\sigma_\text{WBF}\times\mathrm{d}\gamma_b$ were dropped to maintain a strict fixed-order expansion.
Such a strict fixed-order treatment is however incompatible with the use of a parton shower, which resums specific classes of logarithmically-enhanced contributions to all orders in the strong coupling constant $\alpha_s$.
For this reason, in this paper we often work with the best approximation for the decay $\mathrm{d}\gamma_b$, which is a fixed-order description matched to a parton shower, regardless of the approximation used to describe the production process.

Since we know from the analysis of Ref.~\cite{Asteriadis:2024nbg} that treating the Higgs-boson decay at leading order provides a very poor approximation for the fiducial cross section, we consider two distinct descriptions of the $H\to b\bar{b}$ decay: the leading-order approximation supplemented with a parton shower (LO+PS) and an NNLO approximation using the \texttt{MiNLO} method matched to a parton shower (MiNNLO+PS).
The reweighted \texttt{MiNLO} events are generated using \texttt{POWHEG}~\cite{Nason:2004rx,Frixione:2007vw,Alioli:2010xd}, starting with the process $H\to b\bar b g$ at NLO accuracy. The $b$-quarks are treated in the massless approximation. The NLO calculation is then upgraded with the inclusion of the \texttt{MiNLO} Sudakov form factor, subtraction terms and the choice of the coupling such that the resulting cross section is NLO accurate for both $H\to b\bar b g$ and $H\to b\bar b$ processes. A reweighting to the NNLO-accurate $H \to b \bar b$ inclusive width is then sufficient to guarantee the NNLO accuracy of the differential Higgs-boson decay rate in the massless approximation~\cite{Bizon:2019tfo}. Finally, the reweighted events are showered using \texttt{Pythia8.3}~\cite{Bierlich:2022pfr}.

For each of the two approximations to the $H \to b \bar b +X$ decay rate, described in the previous paragraph, we prepare a sample of $10^6$ decay events, generated in the Higgs-boson rest frame, which we use in the partonic integrator for weak-boson fusion.
We modified the Monte-Carlo integrator for this process developed in Refs.~\cite{Asteriadis:2021gpd,Asteriadis:2024nbg} to randomly sample one of these $10^6$ Higgs-boson decay events for each production event.\footnote{Here it is understood that one and the same rest-frame $H \to b \bar b$ event is used for each production event and all counter-events that are required to make it infra-red finite.} We emphasize that the decay events are generated and showered in the Higgs-boson rest frame, but then boosted to the laboratory frame using the information about Higgs-boson kinematics from the production subprocess.
The decay products are then combined with the partons originating from the production subprocess, and clustered using the standard anti-$k_\perp$ jet algorithm~\cite{Cacciari:2008gp}.
The above-mentioned Monte-Carlo simulations of the combined process $pp\to H(b\bar{b})+2j+X$ required approximately $\num{3e5}$ CPU hours to produce the final results that we report below.

We estimate that the use of only a finite number of $10^6$ decay events introduces a sampling error of order $\mathcal{O}(\qty{0.1}{\percent})$, which is negligible in comparison with other uncertainties that we discuss below.
We arrive at this estimate by using two independent samples of decay events and comparing respective results for the fiducial cross sections, defined below.

There are important differences in how $b$-quarks are treated between the two approximations that we use to describe the $H \to b \bar b$ decay.
Both in the leading-order approximation to $H \to b \bar b$ and in the parton shower, the $b$-quarks are taken to be massive, with the mass $m_b=\qty{4.78}{\GeV}$.
On the contrary, the \texttt{MiNLO} implementation of the $H\to b\bar{b}$ decay relies on massless matrix elements. However, \texttt{POWHEG}, which is needed to generate
reweighted \texttt{MiNLO} events, uses the bottom mass to account for the so-called dead-cone effect around the $b$-quarks. Hence, for the compatibility between \texttt{MiNLO}
and \texttt{POWHEG}, the momenta of decay events generated in the massless approximation are ``reshuffled'' to provide a finite mass to $b$-quarks. This bottom-quark mass serves as a collinear regulator in the parton shower.

In the LO+PS approximation the initial scale of the parton shower is set to $m_H/2=\qty{62.5}{\GeV}$.
On the other hand, when events are matched to the parton shower in the \texttt{POWHEG} methods, the starting scale of the shower is set to the transverse momentum of the emitted parton relative to the emitter. This achieves full coverage of the phase space, and avoids double-counting, ensuring that the parton shower does not spoil the accuracy of the fixed-order calculation. Because of the reshuffling procedure used to introduce a $b$-quark mass, there is an ambiguity in the starting scale of the shower, which can be computed either using massless or reshuffled massive momenta. We find that this ambiguity leads to differences of the order of $\qty{0.5}{\percent}$, which are small compared to other uncertainties discussed in the next section. In the following we show results where the starting scale of the shower is computed with the ``massless'' momenta, according to Eq.~(A.1) in Ref.~\cite{Campbell:2014kua}. Finally, we note that top quarks are not included in the calculation, and up, down, strange, and charm quarks are always treated as massless.

Similarly to Ref.~\cite{Asteriadis:2024nbg}, we do not fully account for $b$-jets originating in the production subprocess. Indeed, $b$-quarks in the WBF process are treated as massless, and are not tagged as $b$-flavored for the purpose of $b$-jet identification. This implies that in our simulation the $b$-jets originate \emph{exclusively} from the decay subprocess. The rationale for this approximation, as well as its expected (one percent) accuracy, was discussed in Ref.~\cite{Asteriadis:2024nbg}.

Finally, we note that we do not include any electroweak corrections, which are known to reduce the WBF production cross section by $\mathcal{O}(\qtyrangemath{5}{7}{\percent})$~\cite{Ciccolini:2007ec} and the $H\to b\bar{b}$ decay rate by $\mathcal{O}(\qty{0.5}{\percent})$ \cite{Dabelstein:1991ky,Bardin:1990zj}. Likewise, only QCD radiation is included in the parton shower that we use in this paper.

%%%%%%%%%%%%%%%%%%%%%%%%%%%%%%%%%%%%%%%%%%%%%%%%%%

\section{Results}
\label{sec:results}
\begin{table}
    \caption{
        Fiducial cross sections for the combined process $pp{\to}H(b \bar b){+}2j$ computed using different approximations for the production (rows) and for the decay (columns) subprocesses.
        The renormalization and factorization scales are set to their central values both for the production and decay subprocesses, see text for further details.
        The Monte-Carlo integration uncertainty is shown, unless absolutely negligible. All values are given in femtobarn.
    }
    {\def\arraystretch{1.5}
    \setlength{\tabcolsep}{5pt}
    \begin{tabular}{ccccc}
        \toprule
	    $\sigma/\unit{\femto\barn}$&fixed order~\cite{Asteriadis:2024nbg}&$\mathrm{LO{+}PS}$&$\mathrm{MiNNLO{+}PS}$\\
        \colrule
	    $\mathrm{LO}$&$75.6\hphantom{(1)}$&$46.6\hphantom{(1)}$&$45.2\hphantom{(1)}$\\
	    $\mathrm{NLO}$&$52.4\hphantom{(1)}$&$43.6(1)$&$42.3\hphantom{(1)}$\\
	    $\mathrm{NNLO}$&$44.6(1)$&$43.1(1)$&$41.4(1)$\\
        \botrule
    \end{tabular}}
    \label{tab:cross-sections}
\end{table}

\begin{figure*}
    \centering
    \includegraphics[height=270pt]{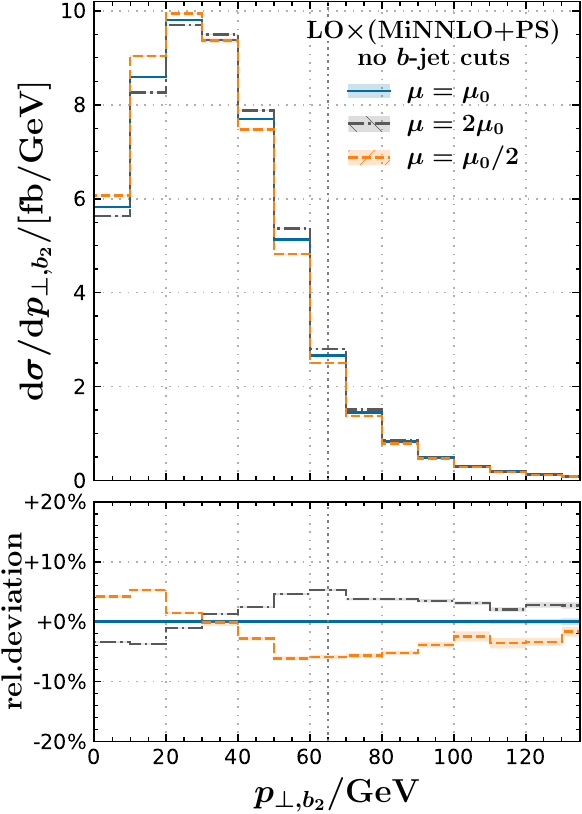}
    \hspace{55pt}
    \includegraphics[height=270pt]{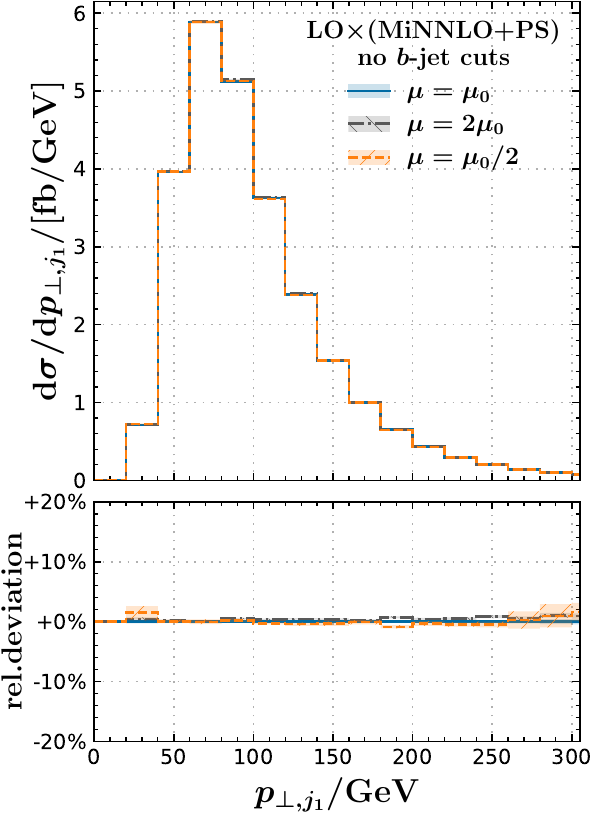}
    \caption{Distributions of transverse momenta
    of the subleading $b$-jet (left) and the leading WBF jet (right) in the approximation $\mathrm{LO{\times}(MiNNLO{+}PS)}$ for different choices of the renormalization scale in the decay subprocess, without the cuts on the transverse momenta and rapidities of the $b$-jets. The bottom panels show relative deviations from the results at the central decay scale $\mu=\mu_0$. The dotted vertical line indicates the position of the $p_{\perp,b_2}$ cut in our default setup, in which only events to the right of the vertical line are accepted. The bands show Monte-Carlo uncertainty only.}
    \label{fig:decscale_no_cuts_pt}
\end{figure*}

\begin{figure*}[t]
    \centering
    \includegraphics[height=270pt]{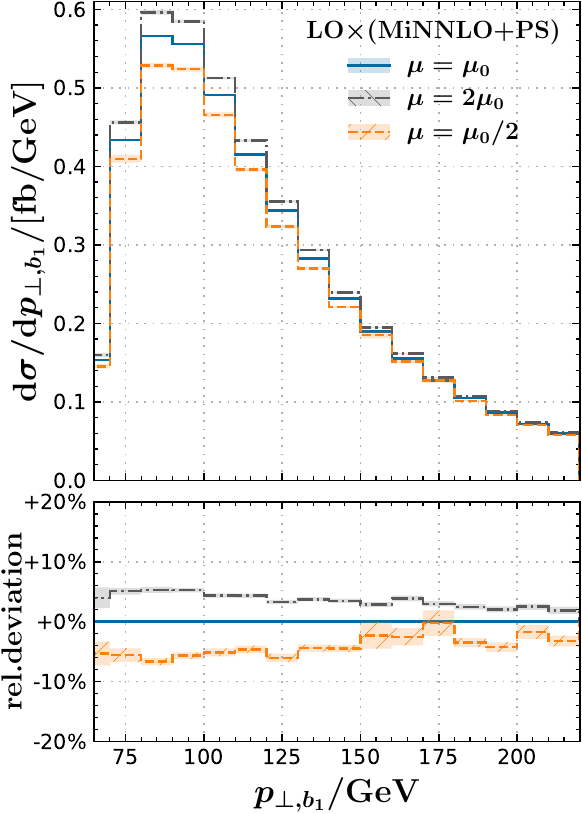}
    \hspace{55pt}
    \includegraphics[height=270pt]{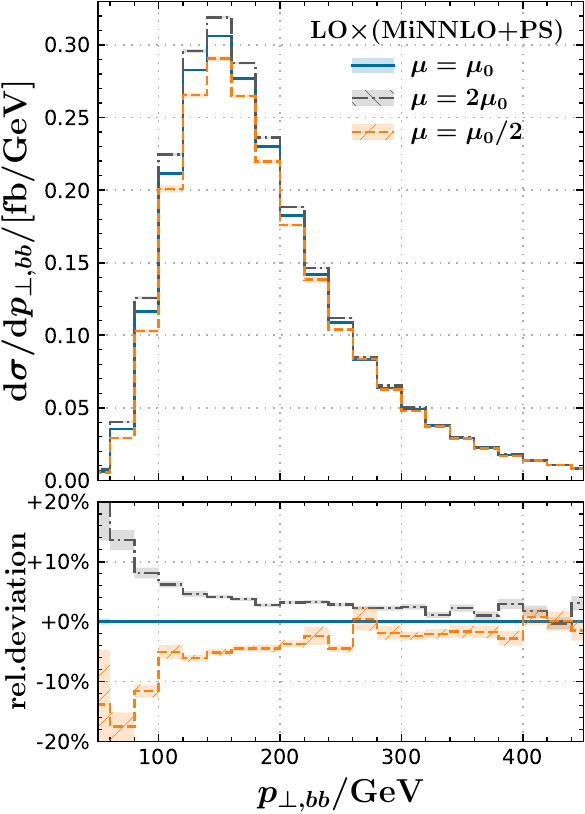}
    \caption{Distributions of the transverse momentum of the leading $b$-jet (left) and the reconstructed Higgs boson (right) in the approximation $\mathrm{LO{\times}(MiNNLO{+}PS)}$ for different choices of the renormalization scale in the decay subprocess, with the default set of $b$-jet cuts. The bottom panels show relative deviations from the results at the central decay scale $\mu=\mu_0$. The bands show Monte-Carlo uncertainty only.}
    \label{fig:decscale_ptb}
\end{figure*}

\begin{figure*}
    \centering
    \includegraphics[height=270pt]{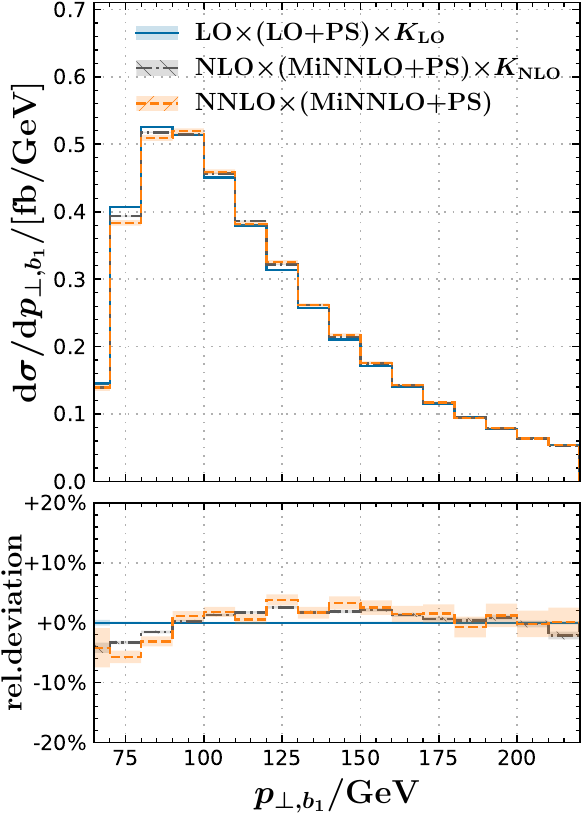}
    \hspace{55pt}
    \includegraphics[height=270pt]{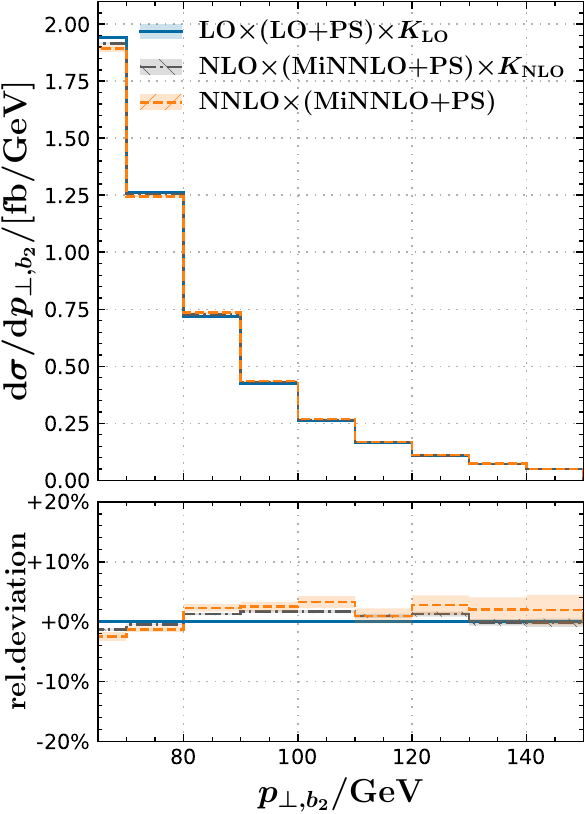}
    \caption{Distributions of transverse momenta of the leading (left) and subleading (right) $b$-jet. \commonPlotAnnotation{}}
    \label{fig:ptb}
\end{figure*}

\begin{figure*}[t]
    \centering
    \includegraphics[height=270pt]{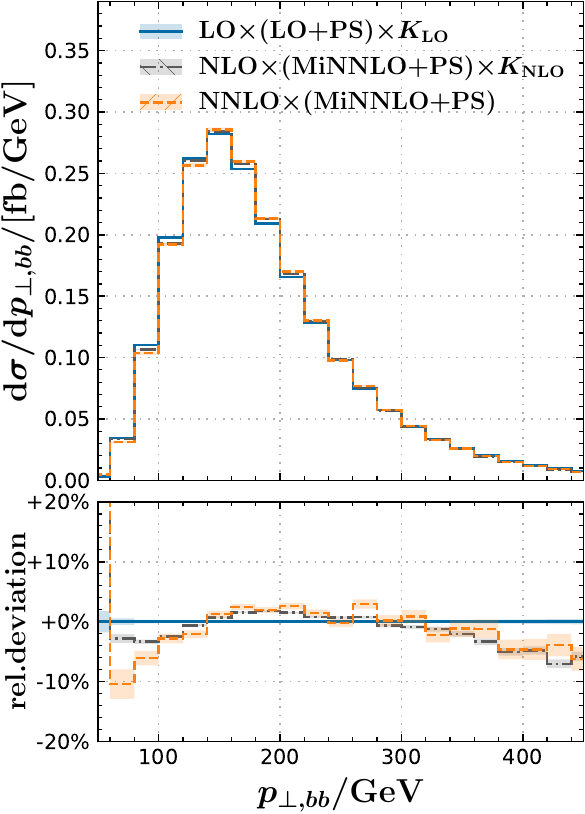}
    \hspace{55pt}
    \includegraphics[height=270pt]{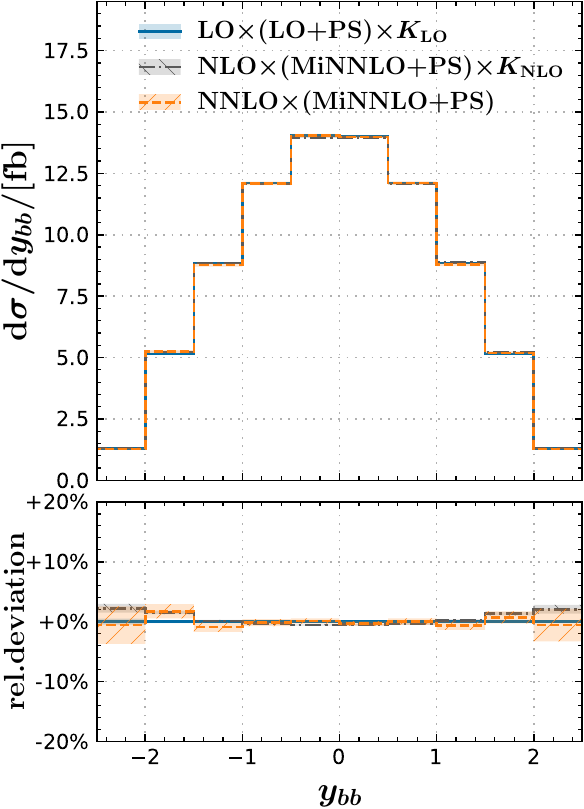}
    \caption{Distributions of transverse momentum of the reconstructed Higgs boson (left) and its rapidity (right). \commonPlotAnnotation{}}
    \label{fig:pt_bb_y_bb}
\end{figure*}

For the numerical modelling of the WBF subprocess we employ the standard setup that has been used earlier in several theoretical calculations~\cite{Cacciari:2015jma,Cruz-Martinez:2018rod,Asteriadis:2021gpd,Asteriadis:2024nbg}.
We consider proton-proton collisions at the center-of-mass energy $E=\qty{13}{\TeV}$.
The mass of the Higgs boson is set to $m_H=\qty{125}{\GeV}$,
the mass of the $W$-boson is $m_W=\qty{80.398}{\GeV}$,
and the mass of the $Z$-boson is $m_Z=\qty{91.1876}{\GeV}$.
Although numerically insignificant, for consistency with the previous calculations
the widths of the $W$- and $Z$-bosons are included in their propagators.
Their numerical values are $\Gamma_W=\qty{2.1054}{\GeV}$ and $\Gamma_Z=\qty{2.4952}{\GeV}$.
All electroweak couplings are derived from the Fermi constant $G_F=\qty{1.6639e-5}{\GeV^{-2}}$.
The Cabibbo-Kobayashi-Maskawa (CKM) matrix is assumed to be equal to the identity matrix; this approximation does not introduce any noticeable
uncertainty thanks to the unitarity of the CKM matrix.

For the WBF simulation the strong coupling constant $\alpha_s$, the parton distribution functions of the proton, and their evolution are taken from the \texttt{LHAPDF}~\cite{Buckley:2014ana} set \texttt{NNPDF31\_nnlo\_as\_0118}~\cite{NNPDF:2017mvq}. The central values of the factorization and renormalization scales are
\begin{equation}
	\mu_F^2=\mu_R^2=\frac{m_H}{2}\sqrt{\frac{m_H^2}{4}+p_{\perp,H}^2},
\end{equation}
where $p_{\perp,H}$ is the transverse momentum of the Higgs boson. We emphasize that these scales refer to the \emph{production} subprocess only and that the choice of the renormalization scale for $H \to b \bar b$ decay is different, as described below.

The event selection criteria used in this study are similar to the ones employed in the ATLAS analysis in Ref.~\cite{ATLAS:2020bhl}, and are identical to the ones in Refs.~\cite{Asteriadis:2021gpd,Asteriadis:2024nbg}.
All jets are defined using the standard anti-$k_\perp$ algorithm~\cite{Cacciari:2008gp} with the jet-radius parameter $R=0.4$.
A jet is considered to be a $b$-jet if it contains at least one bottom quark or antiquark.
In the event analysis the pair of $b$-jets with rapidities $|y_b|<2.5$ and the invariant mass closest to $m_H$ is presumed to originate from Higgs decay.
Both of these $b$-jets are required to have a transverse momentum of $p_{\perp,b}>\qty{65}{\GeV}$. No restrictions are imposed on the transverse momentum $p_{\perp,bb}$ of the selected $b$-jet pair.

In addition to the pair of $b$-jets we require that the event contains a pair of WBF-tagging light jets.
The leading and subleading WBF jets are selected as the two non-$b$-flavored jets with rapidities $|y_j|<4.5$, and the largest and the second-largest transverse momenta $p_{\perp,j_1}, p_{\perp,j_2}$.
Both WBF jets must have transverse momentum $p_{\perp,j}>\qty{25}{\GeV}$, invariant mass $m_{j_1j_2}>\qty{600}{\GeV}$, and a separation in rapidity $|y_{j_1}-y_{j_2}|>4.5$.
The combination of rapidity requirements $|y_j|<4.5$ and $|y_{j_1}-y_{j_2}|>4.5$ implies that the two WBF jets must lie in the opposite hemispheres.

The $H\to b\bar{b}$ subprocess was modelled using the event generator developed in Ref.~\cite{Bizon:2019tfo}.
The value of the $b$-quark Yukawa coupling $\overline{y}_b$ is derived from the Fermi constant $G_F$ and the \MSbar{} $b$-quark mass $\overline{m}_b(\mu=\qty{125}{\GeV})=\qty{2.81}{\GeV}$.
The values of $\overline{m}_b(\mu)$ at other scales $\mu$ are calculated by solving the two-loop evolution equation using \texttt{CRunDec}~\cite{Chetyrkin:2000yt,Schmidt:2012az,Herren:2017osy}.

As already mentioned in Sec.~\ref{sec:details}, the overall normalization of the decay rate cancels between the numerator and the denominator in Eq.~\eqref{eq:gammab} and has no impact on the cross section of the combined WBF process $pp \to H(b\bar{b})+2j$.
Nevertheless, it is instructive to compare the inclusive decay width $\Gamma_{H\to b\bar{b}}^\textrm{MiNNLO}$ calculated using the generator of Ref.~\cite{Bizon:2019tfo} with the results reported in Ref.~\cite{Asteriadis:2024nbg}.
We find
\be
\begin{split}
	\Gamma^\mathrm{NNLO}_{H\to b\bar{b}} & =2.432_{+0.031}^{-0.047}\,\unit{\MeV}\,,
    \\
    \Gamma^\textrm{MiNNLO}_{H\to b\bar{b}} & =2.411_{+0.016}^{-0.039}\,\unit{\MeV}\,,
    \end{split}
    \label{eq5}
\ee
where the sub- and superscripts show the impact of the renormalization-scale variation from the central value by a factor two in both directions (down and up, respectively).
The central renormalization scale is $\mu=m_H$ for the NNLO decay prediction and $\mu=\mu_0=\sqrt{y_3}m_H$ for the MiNNLO result, where $y_3$ is the three-jet resolution parameter in the Cambridge algorithm~\cite{Bizon:2019tfo}.
The two results are compatible within the uncertainty, although the central value of the MiNNLO result is lower by about a percent. This difference is consistent with the fact that the MiNNLO calculation~\cite{Bizon:2019tfo} employs massless $b$-quarks and omits the $y_b y_t$ contributions, that \emph{increase} the NNLO decay width by $\mathcal{O}(0.03)~\unit{\MeV}$.

The fiducial cross sections computed using the different approximations for the WBF production and the $H\to b\bar{b}$ decay subprocesses are shown in Table~\ref{tab:cross-sections}.
One sees that it is possible to use fixed-order description of the decay, extended to \emph{high} perturbative orders, or the parton shower to arrive at fairly consistent results. At the same time, the use of parton shower for the description of the decay already at leading order makes the stability of theoretical predictions manifest, demonstrating that once large effects related to soft radiation are taken into account, the remaining corrections play a minor role. This follows from the comparison of,
e.g., $\mathrm{NNLO}{\times}(\mathrm{LO}{+}\mathrm{PS})$\footnote{Here and in the following, the labels to the left and right of the symbol ``$\times$'' refer respectively to the approximations used for the WBF production and the $H\to b\bar{b}$ decay subprocesses.} approximation with $\mathrm{NNLO}{\times}(\mathrm{MiNNLO}{+}\mathrm{PS})$,
which differ by less than five percent. If, on the other hand, we keep the best description of the decay---which is provided by $\mathrm{MiNNLO}{+}\mathrm{PS}$---and study corrections related to production, the shift from leading order to next-to-next-to-leading order is about ten percent, being fully consistent with earlier fixed-order calculations~\cite{Cacciari:2015jma,Cruz-Martinez:2018rod,Asteriadis:2021gpd} for the production process.

It is conventional to estimate the uncertainty of theoretical predictions for fiducial cross sections by varying the renormalization and factorization scales. This issue was addressed in the fixed-order calculation of Ref.~\cite{Asteriadis:2024nbg}, where it was shown that the scale uncertainty related to the production subprocess is small and amounts to about $\qty{2}{\percent}$ at NNLO.
In the present case the scale sensitivity is even smaller because we do not neglect higher-order terms in the combination of production and decay corrections; which, however, does not imply that the perturbative uncertainty in the production subprocess is less than $\mathcal{O}(\qty{2}{\percent})$.

We will now discuss the dependence of the theoretical prediction on the renormalization scale choice for the decay subprocess. It follows from Eq.~\eqref{eq5} that the scale uncertainties for the total decay width at NNLO are at most two percent. Since they arise, primarily, from the dependence of the Yukawa coupling on the scale choice, and since the Yukawa coupling cancels in the calculation of the fiducial cross section, it is natural to expect that the dependence on the choice of the renormalization scale in the decay is negligible.
However, this expectation is not correct \cite{Asteriadis:2024nbg}; in fact, we find that the residual uncertainty from the choice of the renormalization scale in the decay is about five percent. This (fairly strong) sensitivity to the scale choice in the decay originates from changes to kinematic distributions due to radiation, in the presence of a high cut on the transverse momentum distribution of $b$-jets.

To illustrate this effect, on the left-hand side of Fig.~\ref{fig:decscale_no_cuts_pt} we show the transverse-momentum distributions of the subleading $b$-jet in a setup where restrictions on transverse momenta and rapidities of $b$-jets are lifted.
We observe that corrections have a different sign above and below the peak. In fact, from the sensitivity of the inclusive width to the scale choice shown in Eq.~\eqref{eq5}, we know that they largely compensate each other.
However, they do shift the peaks of the $b$-jet transverse momentum distributions, which has a noticeable impact on the acceptance rate with the $b$-jet cuts. Interestingly,
these scale variations have a negligible impact on the rapidity and light-jet distributions, as illustrated on the right side of Fig.~\ref{fig:decscale_no_cuts_pt}.

Figure~\ref{fig:decscale_ptb} shows the impact of these decay-scale variations on the $b$-jet distributions with our default event selection criteria. We take the resulting changes in the predictions as an estimate of the perturbative uncertainty in the description of the decay subprocess.
This uncertainty is of order $\mathcal{O}(\qty{5}{\percent})$ and, to a large extent, is uniform over the entire kinematic range, except the low-$p_{\perp,bb}$ region $p_{\perp,bb}\lesssim\qty{100}{\GeV}$, where the uncertainty rises to $\mathcal{O}(\qty{10}{\percent})$.

%%%%%%%%%%%%%%%%%%%%%%%%%%%%%%%%%%%%%%%%%%%%%%%%%%

As the next step we consider the impact of QCD corrections on kinematic distributions with the central choice of the renormalization and factorization scales. We note that significant modifications of shapes of such distributions were observed both in the pure production case~\cite{Cacciari:2015jma,Cruz-Martinez:2018rod} and
also when corrections to the decay in the presence of the aggressive $p_\perp$ cut on $b$-jets are taken into account \cite{Asteriadis:2024nbg}. However, since
the use of the parton shower in leading-order computations leads to stabilization of the perturbative expansion, and
since it is rather inexpensive to obtain results using the $\mathrm{LO}{\times}(\mathrm{LO}{+}\mathrm{PS})$ approximation, we decided to compare \emph{shapes} of kinematic distributions calculated using this approximation with the ones computed with the $\mathrm{NNLO}{\times}(\mathrm{MiNNLO}{+}\mathrm{PS})$
setup. Therefore, we rescale the $\mathrm{LO} {\times}(\mathrm{LO}{+}\mathrm{PS})$ and $\mathrm{NLO}{\times}(\mathrm{MiNNLO}{+}\mathrm{PS})$ distributions by $K$ factors such that their integral corresponds to the $\mathrm{NNLO}{\times}(\mathrm{MiNNLO+PS})$ fiducial cross section.

There are two types of distributions that we will consider: distributions that are mostly determined by the dynamics of $H \to b \bar b$ decay, and distributions that are mostly determined by the production process. We may hope that the shapes of distributions of the first type are
well-described in the $\mathrm{LO}{\times}( \mathrm{LO}{+}\mathrm{PS})$ approximation, while distributions of the second type should receive important (but not dramatic) corrections, consistent with modifications observed in the studies of the production process alone~\cite{Cacciari:2015jma,Cruz-Martinez:2018rod,Asteriadis:2021gpd}.

We begin with the distributions of the first type.
The QCD corrections to the transverse momentum distributions of the leading and next-to-leading $b$-jets are shown in Fig.~\ref{fig:ptb}.
As seen from the lower panes of the two figures there, the shapes of both distributions are well-captured by the $\mathrm{LO}{\times}(\mathrm{LO}{+}\mathrm{PS})$
approximation with deviations at a few-percent level only. At the same time, a small shape change in $p_{\perp,b_1}$ distribution is visible, with the reduction of events at $p_{\perp,b_1} \sim \qty{70}{\GeV}$
and a slight increase at around $p_{\perp,b_1} \sim \qty{125}{\GeV}$. Similarly, very small changes in the shape of the transverse momentum distribution of the reconstructed Higgs boson, $p_{\perp,bb}$, are observed in Fig.~\ref{fig:pt_bb_y_bb},
whereas the shape of its rapidity distribution remains unchanged even at a few-percent level, except for large rapidities. These corrections to the distributions of the reconstructed Higgs bosons closely follow corrections to distributions of Higgs bosons produced in simulation of WBF without the decay subprocess~\cite{Asteriadis:2021gpd}.

\begin{figure*}
    \centering
    \includegraphics[height=270pt]{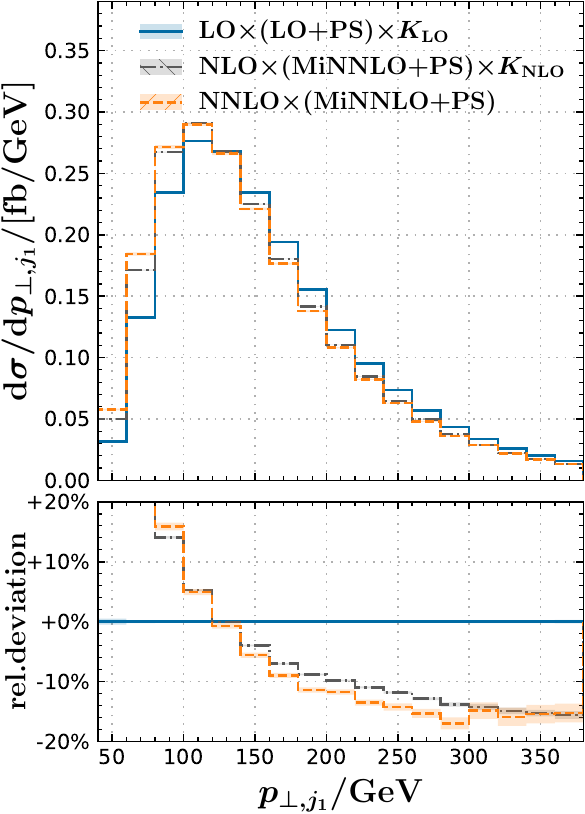}
    \hspace{55pt}
    \includegraphics[height=270pt]{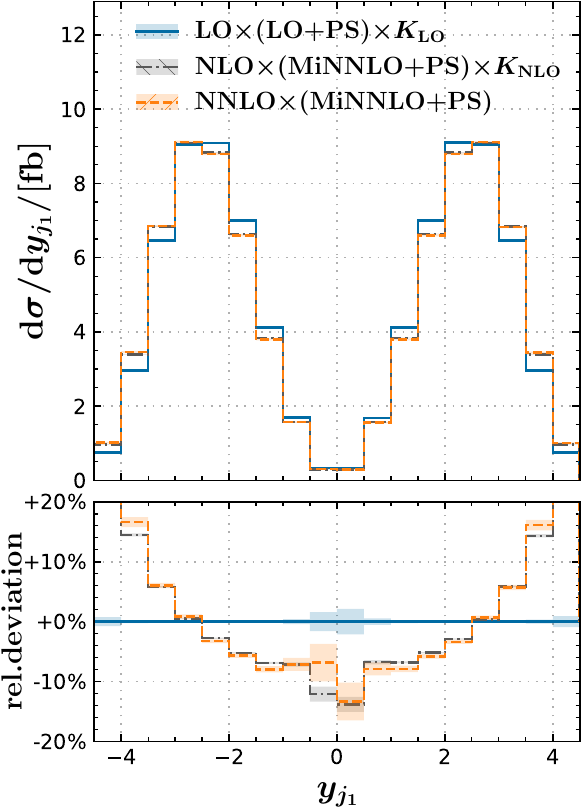}
    \caption{Distributions of the transverse momentum (left)
    and rapidity
    (right) of the leading WBF-tagging jet. \commonPlotAnnotation{}}
    \label{fig:ptj1_yj1}
\end{figure*}

Kinematic distributions of the second type include those that are mostly determined by the production subprocess. The examples are transverse momentum and rapidity distributions of light WBF-tagging jets, which are illustrated in Fig.~\ref{fig:ptj1_yj1}.
As can be seen in these figures, the $\mathrm{LO}{\times}(\mathrm{LO}{+}\mathrm{PS}) $ setup provides a poor approximation in this case, and changes in shapes of the two distributions are consistent with expectations based on fixed-order calculations without the Higgs decay~\cite{Cacciari:2015jma,Cruz-Martinez:2018rod,Asteriadis:2021gpd}, in particular, the effect of higher-order corrections is to render the jets softer.

\section{Conclusions}
\label{sec:conclusions}

In this paper, we have studied QCD effects in Higgs production in weak-boson fusion followed by the decay of the Higgs boson into $b$-jets, focusing on a situation where $b$-jets are selected using an aggressive high-$p_\perp$ cut \cite{ATLAS:2020bhl}.
Earlier, it was observed in Ref.~\cite{Asteriadis:2024nbg} that fiducial cross sections in such a case receive very significant QCD corrections that reduce the leading-order cross section by almost a factor of two. It was conjectured in Ref.~\cite{Asteriadis:2024nbg} that these large QCD effects are connected to soft and collinear QCD radiation and, thus, can be efficiently dealt with by incorporating parton-shower description of $H \to b \bar b$ decay into a ``leading-order'' prediction.
The goal of the current paper is to explicitly check this conjecture using fixed-order description of $H \to b \bar b$ matched to parton shower within the \texttt{MiNLO} framework \cite{Hamilton:2012np,Hamilton:2012rf}.

We have found that if a parton-shower description of $H \to b \bar b$ decay is combined with fixed-order description of Higgs production in WBF, the stability of the perturbative expansion of the combined process improves dramatically. Furthermore, this improvement occurs independently of which parton-shower-matched approximation is used for the decay ($\mathrm{LO}{+}\mathrm{PS}$ vs. $\mathrm{MiNNLO}{+}\mathrm{PS}$). The change in the fiducial cross section from the (modified) leading-order prediction to the NNLO one, is about ten percent, which should be compared with $\mathcal{O}(\qty{40}{\percent})$ if only fixed-order computations are used. Similarly, shapes of kinematic distributions \emph{driven by the decay dynamics} are well-described by parton showers with only relatively minor modifications from matching to fixed orders.
%%%%%%%%%%%%%%%%%%%%%%%%%%%%%%%%%%%%%%%%%%%%%%%%%%

An important question is the remaining uncertainty of the theoretical prediction for the $pp \to H(b \bar b) + 2j$ process, with a relatively high cut on the $b$-jet transverse momentum.
To estimate it, we can consider the various results in the last row in Table~\ref{tab:cross-sections}. The spread of these predictions is about five percent if one excludes the NNLO fixed-order result. In addition, there is a two-percent scale uncertainty related to the production subprocess and about five-percent uncertainty related to the choice of the renormalization scale in the decay process.
As we mentioned earlier, this uncertainty does not affect shapes of the distributions in the fiducial region.

All in all one should associate a $\mathcal{O}(\qtyrangemath{5}{7}{\percent})$ uncertainty with the $\mathrm{NNLO}{\times}(\mathrm{MiNNLO}{+}\mathrm{PS})$ estimate of the fiducial cross section in Table~\ref{tab:cross-sections}.
To reduce it further, one could perhaps make use of N$^3$LO QCD computation of $H \to b \bar b$ decay reported in Ref.~\cite{Mondini:2019gid}, although methods to combine perturbative computations of such a high order with parton showers do not exist yet.
Given the improved uncertainty estimate of the fiducial cross section, the electroweak corrections to WBF, which are known to reduce the cross section by $\mathcal{O}(\qtyrangemath{5}{7}{\percent})$~\cite{Ciccolini:2007ec}, need to be accounted for.
Furthermore, other so-far neglected effects, such as $b$-jets arising from the WBF production process, will have to be investigated.
Finally, a full matching including parton-shower effects in production---using the \texttt{MiNNLO} procedure on each fermion line---would be useful for a better modelling of the tagging jets.

\begin{acknowledgments}
We are grateful to R.~R\"ontsch for useful conversations. The research of KM and IN is partially supported by the Deutsche Forschungsgemeinschaft (DFG, German Research Foundation) under grant 396021762 - TRR 257.
\end{acknowledgments}

\bibliography{vbfhbb}

%apsrev4-2.bst 2019-01-14 (MD) hand-edited version of apsrev4-1.bst
%Control: key (0)
%Control: author (8) initials jnrlst
%Control: editor formatted (1) identically to author
%Control: production of article title (0) allowed
%Control: page (0) single
%Control: year (1) truncated
%Control: production of eprint (0) enabled
\begin{thebibliography}{26}%
\makeatletter
\providecommand \@ifxundefined [1]{%
 \@ifx{#1\undefined}
}%
\providecommand \@ifnum [1]{%
 \ifnum #1\expandafter \@firstoftwo
 \else \expandafter \@secondoftwo
 \fi
}%
\providecommand \@ifx [1]{%
 \ifx #1\expandafter \@firstoftwo
 \else \expandafter \@secondoftwo
 \fi
}%
\providecommand \natexlab [1]{#1}%
\providecommand \enquote  [1]{``#1''}%
\providecommand \bibnamefont  [1]{#1}%
\providecommand \bibfnamefont [1]{#1}%
\providecommand \citenamefont [1]{#1}%
\providecommand \href@noop [0]{\@secondoftwo}%
\providecommand \href [0]{\begingroup \@sanitize@url \@href}%
\providecommand \@href[1]{\@@startlink{#1}\@@href}%
\providecommand \@@href[1]{\endgroup#1\@@endlink}%
\providecommand \@sanitize@url [0]{\catcode `\\12\catcode `\$12\catcode
  `\&12\catcode `\#12\catcode `\^12\catcode `\_12\catcode `\%12\relax}%
\providecommand \@@startlink[1]{}%
\providecommand \@@endlink[0]{}%
\providecommand \url  [0]{\begingroup\@sanitize@url \@url }%
\providecommand \@url [1]{\endgroup\@href {#1}{\urlprefix }}%
\providecommand \urlprefix  [0]{URL }%
\providecommand \Eprint [0]{\href }%
\providecommand \doibase [0]{https://doi.org/}%
\providecommand \selectlanguage [0]{\@gobble}%
\providecommand \bibinfo  [0]{\@secondoftwo}%
\providecommand \bibfield  [0]{\@secondoftwo}%
\providecommand \translation [1]{[#1]}%
\providecommand \BibitemOpen [0]{}%
\providecommand \bibitemStop [0]{}%
\providecommand \bibitemNoStop [0]{.\EOS\space}%
\providecommand \EOS [0]{\spacefactor3000\relax}%
\providecommand \BibitemShut  [1]{\csname bibitem#1\endcsname}%
\let\auto@bib@innerbib\@empty
%</preamble>
\bibitem [{\citenamefont {Asteriadis}\ \emph {et~al.}(2024)\citenamefont
  {Asteriadis}, \citenamefont {Behring}, \citenamefont {Melnikov},
  \citenamefont {Novikov},\ and\ \citenamefont
  {R\"ontsch}}]{Asteriadis:2024nbg}%
  \BibitemOpen
  \bibfield  {author} {\bibinfo {author} {\bibfnamefont {K.}~\bibnamefont
  {Asteriadis}}, \bibinfo {author} {\bibfnamefont {A.}~\bibnamefont {Behring}},
  \bibinfo {author} {\bibfnamefont {K.}~\bibnamefont {Melnikov}}, \bibinfo
  {author} {\bibfnamefont {I.}~\bibnamefont {Novikov}},\ and\ \bibinfo {author}
  {\bibfnamefont {R.}~\bibnamefont {R\"ontsch}},\ }\bibfield  {title} {\bibinfo
  {title} {{QCD corrections to Higgs boson production and $H\to b\bar{b}$ decay
  in weak boson fusion}},\ }\href {https://doi.org/10.1103/PhysRevD.110.054017}
  {\bibfield  {journal} {\bibinfo  {journal} {Phys. Rev. D}\ }\textbf {\bibinfo
  {volume} {110}},\ \bibinfo {pages} {054017} (\bibinfo {year} {2024})},\
  \Eprint {https://arxiv.org/abs/2407.09363} {arXiv:2407.09363 [hep-ph]}
  \BibitemShut {NoStop}%
\bibitem [{\citenamefont {Aad}\ \emph {et~al.}(2022)\citenamefont {Aad} \emph
  {et~al.}}]{ATLAS:2022vkf}%
  \BibitemOpen
  \bibfield  {author} {\bibinfo {author} {\bibfnamefont {G.}~\bibnamefont
  {Aad}} \emph {et~al.} (\bibinfo {collaboration} {ATLAS}),\ }\bibfield
  {title} {\bibinfo {title} {{A detailed map of Higgs boson interactions by the
  ATLAS experiment ten years after the discovery}},\ }\href
  {https://doi.org/10.1038/s41586-022-04893-w} {\bibfield  {journal} {\bibinfo
  {journal} {Nature}\ }\textbf {\bibinfo {volume} {607}},\ \bibinfo {pages}
  {52} (\bibinfo {year} {2022})},\ \bibinfo {note} {[Erratum: Nature 612, E24
  (2022)]},\ \Eprint {https://arxiv.org/abs/2207.00092} {arXiv:2207.00092
  [hep-ex]} \BibitemShut {NoStop}%
\bibitem [{\citenamefont {Tumasyan}\ \emph {et~al.}(2022)\citenamefont
  {Tumasyan} \emph {et~al.}}]{CMS:2022dwd}%
  \BibitemOpen
  \bibfield  {author} {\bibinfo {author} {\bibfnamefont {A.}~\bibnamefont
  {Tumasyan}} \emph {et~al.} (\bibinfo {collaboration} {CMS}),\ }\bibfield
  {title} {\bibinfo {title} {{A portrait of the Higgs boson by the CMS
  experiment ten years after the discovery.}},\ }\href
  {https://doi.org/10.1038/s41586-022-04892-x} {\bibfield  {journal} {\bibinfo
  {journal} {Nature}\ }\textbf {\bibinfo {volume} {607}},\ \bibinfo {pages}
  {60} (\bibinfo {year} {2022})},\ \bibinfo {note} {[Erratum: Nature 623, E4
  (2023)]},\ \Eprint {https://arxiv.org/abs/2207.00043} {arXiv:2207.00043
  [hep-ex]} \BibitemShut {NoStop}%
\bibitem [{\citenamefont {Aad}\ \emph {et~al.}(2021)\citenamefont {Aad} \emph
  {et~al.}}]{ATLAS:2020bhl}%
  \BibitemOpen
  \bibfield  {author} {\bibinfo {author} {\bibfnamefont {G.}~\bibnamefont
  {Aad}} \emph {et~al.} (\bibinfo {collaboration} {ATLAS}),\ }\bibfield
  {title} {\bibinfo {title} {{Measurements of Higgs bosons decaying to bottom
  quarks from vector boson fusion production with the ATLAS experiment at
  $\sqrt{s}=13\,\text {TeV}$}},\ }\href
  {https://doi.org/10.1140/epjc/s10052-021-09192-8} {\bibfield  {journal}
  {\bibinfo  {journal} {Eur. Phys. J. C}\ }\textbf {\bibinfo {volume} {81}},\
  \bibinfo {pages} {537} (\bibinfo {year} {2021})},\ \Eprint
  {https://arxiv.org/abs/2011.08280} {arXiv:2011.08280 [hep-ex]} \BibitemShut
  {NoStop}%
\bibitem [{\citenamefont {Cacciari}\ \emph {et~al.}(2015)\citenamefont
  {Cacciari}, \citenamefont {Dreyer}, \citenamefont {Karlberg}, \citenamefont
  {Salam},\ and\ \citenamefont {Zanderighi}}]{Cacciari:2015jma}%
  \BibitemOpen
  \bibfield  {author} {\bibinfo {author} {\bibfnamefont {M.}~\bibnamefont
  {Cacciari}}, \bibinfo {author} {\bibfnamefont {F.~A.}\ \bibnamefont
  {Dreyer}}, \bibinfo {author} {\bibfnamefont {A.}~\bibnamefont {Karlberg}},
  \bibinfo {author} {\bibfnamefont {G.~P.}\ \bibnamefont {Salam}},\ and\
  \bibinfo {author} {\bibfnamefont {G.}~\bibnamefont {Zanderighi}},\ }\bibfield
   {title} {\bibinfo {title} {{Fully Differential Vector-Boson-Fusion Higgs
  Production at Next-to-Next-to-Leading Order}},\ }\href
  {https://doi.org/10.1103/PhysRevLett.115.082002} {\bibfield  {journal}
  {\bibinfo  {journal} {Phys. Rev. Lett.}\ }\textbf {\bibinfo {volume} {115}},\
  \bibinfo {pages} {082002} (\bibinfo {year} {2015})},\ \bibinfo {note}
  {[Erratum: Phys.Rev.Lett. 120, 139901(E) (2018)]},\ \Eprint
  {https://arxiv.org/abs/1506.02660} {arXiv:1506.02660 [hep-ph]} \BibitemShut
  {NoStop}%
\bibitem [{\citenamefont {Cruz-Martinez}\ \emph {et~al.}(2018)\citenamefont
  {Cruz-Martinez}, \citenamefont {Gehrmann}, \citenamefont {Glover},\ and\
  \citenamefont {Huss}}]{Cruz-Martinez:2018rod}%
  \BibitemOpen
  \bibfield  {author} {\bibinfo {author} {\bibfnamefont {J.}~\bibnamefont
  {Cruz-Martinez}}, \bibinfo {author} {\bibfnamefont {T.}~\bibnamefont
  {Gehrmann}}, \bibinfo {author} {\bibfnamefont {E.~W.~N.}\ \bibnamefont
  {Glover}},\ and\ \bibinfo {author} {\bibfnamefont {A.}~\bibnamefont {Huss}},\
  }\bibfield  {title} {\bibinfo {title} {{Second-order QCD effects in Higgs
  boson production through vector boson fusion}},\ }\href
  {https://doi.org/10.1016/j.physletb.2018.04.046} {\bibfield  {journal}
  {\bibinfo  {journal} {Phys. Lett. B}\ }\textbf {\bibinfo {volume} {781}},\
  \bibinfo {pages} {672} (\bibinfo {year} {2018})},\ \Eprint
  {https://arxiv.org/abs/1802.02445} {arXiv:1802.02445 [hep-ph]} \BibitemShut
  {NoStop}%
\bibitem [{\citenamefont {Asteriadis}\ \emph {et~al.}(2022)\citenamefont
  {Asteriadis}, \citenamefont {Caola}, \citenamefont {Melnikov},\ and\
  \citenamefont {R\"ontsch}}]{Asteriadis:2021gpd}%
  \BibitemOpen
  \bibfield  {author} {\bibinfo {author} {\bibfnamefont {K.}~\bibnamefont
  {Asteriadis}}, \bibinfo {author} {\bibfnamefont {F.}~\bibnamefont {Caola}},
  \bibinfo {author} {\bibfnamefont {K.}~\bibnamefont {Melnikov}},\ and\
  \bibinfo {author} {\bibfnamefont {R.}~\bibnamefont {R\"ontsch}},\ }\bibfield
  {title} {\bibinfo {title} {{NNLO QCD corrections to weak boson fusion Higgs
  boson production in the H \textrightarrow{} b$ \overline{b} $ and H
  \textrightarrow{} WW$^{*}$ \textrightarrow{} 4l decay channels}},\ }\href
  {https://doi.org/10.1007/JHEP02(2022)046} {\bibfield  {journal} {\bibinfo
  {journal} {{JHEP}}\ }\textbf {\bibinfo {volume} {02}},\ \bibinfo {pages}
  {046} (\bibinfo {year} {2022})},\ \Eprint {https://arxiv.org/abs/2110.02818}
  {arXiv:2110.02818 [hep-ph]} \BibitemShut {NoStop}%
\bibitem [{\citenamefont {Bizo\'n}\ \emph {et~al.}(2020)\citenamefont
  {Bizo\'n}, \citenamefont {Re},\ and\ \citenamefont
  {Zanderighi}}]{Bizon:2019tfo}%
  \BibitemOpen
  \bibfield  {author} {\bibinfo {author} {\bibfnamefont {W.}~\bibnamefont
  {Bizo\'n}}, \bibinfo {author} {\bibfnamefont {E.}~\bibnamefont {Re}},\ and\
  \bibinfo {author} {\bibfnamefont {G.}~\bibnamefont {Zanderighi}},\ }\bibfield
   {title} {\bibinfo {title} {{NNLOPS description of the $H \to b\overline{b} $
  decay with MiNLO}},\ }\href {https://doi.org/10.1007/JHEP06(2020)006}
  {\bibfield  {journal} {\bibinfo  {journal} {{JHEP}}\ }\textbf {\bibinfo
  {volume} {06}},\ \bibinfo {pages} {006} (\bibinfo {year} {2020})},\ \Eprint
  {https://arxiv.org/abs/1912.09982} {arXiv:1912.09982 [hep-ph]} \BibitemShut
  {NoStop}%
\bibitem [{\citenamefont {Bierlich}\ \emph {et~al.}(2022)\citenamefont
  {Bierlich} \emph {et~al.}}]{Bierlich:2022pfr}%
  \BibitemOpen
  \bibfield  {author} {\bibinfo {author} {\bibfnamefont {C.}~\bibnamefont
  {Bierlich}} \emph {et~al.},\ }\bibfield  {title} {\bibinfo {title} {{A
  comprehensive guide to the physics and usage of PYTHIA 8.3}},\ }\href
  {https://doi.org/10.21468/SciPostPhysCodeb.8} {\bibfield  {journal} {\bibinfo
   {journal} {SciPost Phys. Codeb.}\ }\textbf {\bibinfo {volume} {2022}},\
  \bibinfo {pages} {8} (\bibinfo {year} {2022})},\ \Eprint
  {https://arxiv.org/abs/2203.11601} {arXiv:2203.11601 [hep-ph]} \BibitemShut
  {NoStop}%
\bibitem [{\citenamefont {Nason}(2004)}]{Nason:2004rx}%
  \BibitemOpen
  \bibfield  {author} {\bibinfo {author} {\bibfnamefont {P.}~\bibnamefont
  {Nason}},\ }\bibfield  {title} {\bibinfo {title} {{A New method for combining
  NLO QCD with shower Monte Carlo algorithms}},\ }\href
  {https://doi.org/10.1088/1126-6708/2004/11/040} {\bibfield  {journal}
  {\bibinfo  {journal} {JHEP}\ }\textbf {\bibinfo {volume} {11}},\ \bibinfo
  {pages} {040}},\ \Eprint {https://arxiv.org/abs/hep-ph/0409146}
  {arXiv:hep-ph/0409146} \BibitemShut {NoStop}%
\bibitem [{\citenamefont {Frixione}\ \emph {et~al.}(2007)\citenamefont
  {Frixione}, \citenamefont {Nason},\ and\ \citenamefont
  {Oleari}}]{Frixione:2007vw}%
  \BibitemOpen
  \bibfield  {author} {\bibinfo {author} {\bibfnamefont {S.}~\bibnamefont
  {Frixione}}, \bibinfo {author} {\bibfnamefont {P.}~\bibnamefont {Nason}},\
  and\ \bibinfo {author} {\bibfnamefont {C.}~\bibnamefont {Oleari}},\
  }\bibfield  {title} {\bibinfo {title} {{Matching NLO QCD computations with
  Parton Shower simulations: the POWHEG method}},\ }\href
  {https://doi.org/10.1088/1126-6708/2007/11/070} {\bibfield  {journal}
  {\bibinfo  {journal} {JHEP}\ }\textbf {\bibinfo {volume} {11}},\ \bibinfo
  {pages} {070}},\ \Eprint {https://arxiv.org/abs/0709.2092} {arXiv:0709.2092
  [hep-ph]} \BibitemShut {NoStop}%
\bibitem [{\citenamefont {Alioli}\ \emph {et~al.}(2010)\citenamefont {Alioli},
  \citenamefont {Nason}, \citenamefont {Oleari},\ and\ \citenamefont
  {Re}}]{Alioli:2010xd}%
  \BibitemOpen
  \bibfield  {author} {\bibinfo {author} {\bibfnamefont {S.}~\bibnamefont
  {Alioli}}, \bibinfo {author} {\bibfnamefont {P.}~\bibnamefont {Nason}},
  \bibinfo {author} {\bibfnamefont {C.}~\bibnamefont {Oleari}},\ and\ \bibinfo
  {author} {\bibfnamefont {E.}~\bibnamefont {Re}},\ }\bibfield  {title}
  {\bibinfo {title} {{A general framework for implementing NLO calculations in
  shower Monte Carlo programs: the POWHEG BOX}},\ }\href
  {https://doi.org/10.1007/JHEP06(2010)043} {\bibfield  {journal} {\bibinfo
  {journal} {JHEP}\ }\textbf {\bibinfo {volume} {06}},\ \bibinfo {pages}
  {043}},\ \Eprint {https://arxiv.org/abs/1002.2581} {arXiv:1002.2581 [hep-ph]}
  \BibitemShut {NoStop}%
\bibitem [{\citenamefont {Hamilton}\ \emph {et~al.}(2012)\citenamefont
  {Hamilton}, \citenamefont {Nason},\ and\ \citenamefont
  {Zanderighi}}]{Hamilton:2012np}%
  \BibitemOpen
  \bibfield  {author} {\bibinfo {author} {\bibfnamefont {K.}~\bibnamefont
  {Hamilton}}, \bibinfo {author} {\bibfnamefont {P.}~\bibnamefont {Nason}},\
  and\ \bibinfo {author} {\bibfnamefont {G.}~\bibnamefont {Zanderighi}},\
  }\bibfield  {title} {\bibinfo {title} {{MINLO: Multi-Scale Improved NLO}},\
  }\href {https://doi.org/10.1007/JHEP10(2012)155} {\bibfield  {journal}
  {\bibinfo  {journal} {JHEP}\ }\textbf {\bibinfo {volume} {10}},\ \bibinfo
  {pages} {155}},\ \Eprint {https://arxiv.org/abs/1206.3572} {arXiv:1206.3572
  [hep-ph]} \BibitemShut {NoStop}%
\bibitem [{\citenamefont {Hamilton}\ \emph {et~al.}(2013)\citenamefont
  {Hamilton}, \citenamefont {Nason}, \citenamefont {Oleari},\ and\
  \citenamefont {Zanderighi}}]{Hamilton:2012rf}%
  \BibitemOpen
  \bibfield  {author} {\bibinfo {author} {\bibfnamefont {K.}~\bibnamefont
  {Hamilton}}, \bibinfo {author} {\bibfnamefont {P.}~\bibnamefont {Nason}},
  \bibinfo {author} {\bibfnamefont {C.}~\bibnamefont {Oleari}},\ and\ \bibinfo
  {author} {\bibfnamefont {G.}~\bibnamefont {Zanderighi}},\ }\bibfield  {title}
  {\bibinfo {title} {{Merging H/W/Z + 0 and 1 jet at NLO with no merging scale:
  a path to parton shower + NNLO matching}},\ }\href
  {https://doi.org/10.1007/JHEP05(2013)082} {\bibfield  {journal} {\bibinfo
  {journal} {JHEP}\ }\textbf {\bibinfo {volume} {05}},\ \bibinfo {pages}
  {082}},\ \Eprint {https://arxiv.org/abs/1212.4504} {arXiv:1212.4504 [hep-ph]}
  \BibitemShut {NoStop}%
\bibitem [{\citenamefont {de~Florian}\ \emph {et~al.}(2016)\citenamefont
  {de~Florian} \emph {et~al.}}]{LHCHiggsCrossSectionWorkingGroup:2016ypw}%
  \BibitemOpen
  \bibfield  {author} {\bibinfo {author} {\bibfnamefont {D.}~\bibnamefont
  {de~Florian}} \emph {et~al.} (\bibinfo {collaboration} {LHC Higgs Cross
  Section Working Group}),\ }\href {https://doi.org/10.23731/CYRM-2017-002}
  {\emph {\bibinfo {title} {{Handbook of LHC Higgs Cross Sections: 4.
  Deciphering the Nature of the Higgs Sector}}}},\ \bibinfo {series} {CERN
  Yellow Reports: Monographs}, Vol.\ \bibinfo {volume} {2/2017}\ (\bibinfo
  {publisher} {CERN},\ \bibinfo {year} {2016})\ \Eprint
  {https://arxiv.org/abs/1610.07922} {arXiv:1610.07922 [hep-ph]} \BibitemShut
  {NoStop}%
\bibitem [{\citenamefont {Cacciari}\ \emph {et~al.}(2008)\citenamefont
  {Cacciari}, \citenamefont {Salam},\ and\ \citenamefont
  {Soyez}}]{Cacciari:2008gp}%
  \BibitemOpen
  \bibfield  {author} {\bibinfo {author} {\bibfnamefont {M.}~\bibnamefont
  {Cacciari}}, \bibinfo {author} {\bibfnamefont {G.~P.}\ \bibnamefont
  {Salam}},\ and\ \bibinfo {author} {\bibfnamefont {G.}~\bibnamefont {Soyez}},\
  }\bibfield  {title} {\bibinfo {title} {{The anti-$k_t$ jet clustering
  algorithm}},\ }\href {https://doi.org/10.1088/1126-6708/2008/04/063}
  {\bibfield  {journal} {\bibinfo  {journal} {{JHEP}}\ }\textbf {\bibinfo
  {volume} {04}},\ \bibinfo {pages} {063} (\bibinfo {year} {2008})},\ \Eprint
  {https://arxiv.org/abs/0802.1189} {arXiv:0802.1189 [hep-ph]} \BibitemShut
  {NoStop}%
\bibitem [{\citenamefont {Campbell}\ \emph {et~al.}(2015)\citenamefont
  {Campbell}, \citenamefont {Ellis}, \citenamefont {Nason},\ and\ \citenamefont
  {Re}}]{Campbell:2014kua}%
  \BibitemOpen
  \bibfield  {author} {\bibinfo {author} {\bibfnamefont {J.~M.}\ \bibnamefont
  {Campbell}}, \bibinfo {author} {\bibfnamefont {R.~K.}\ \bibnamefont {Ellis}},
  \bibinfo {author} {\bibfnamefont {P.}~\bibnamefont {Nason}},\ and\ \bibinfo
  {author} {\bibfnamefont {E.}~\bibnamefont {Re}},\ }\bibfield  {title}
  {\bibinfo {title} {{Top-Pair Production and Decay at NLO Matched with Parton
  Showers}},\ }\href {https://doi.org/10.1007/JHEP04(2015)114} {\bibfield
  {journal} {\bibinfo  {journal} {JHEP}\ }\textbf {\bibinfo {volume} {04}},\
  \bibinfo {pages} {114}},\ \Eprint {https://arxiv.org/abs/1412.1828}
  {arXiv:1412.1828 [hep-ph]} \BibitemShut {NoStop}%
\bibitem [{\citenamefont {Ciccolini}\ \emph {et~al.}(2008)\citenamefont
  {Ciccolini}, \citenamefont {Denner},\ and\ \citenamefont
  {Dittmaier}}]{Ciccolini:2007ec}%
  \BibitemOpen
  \bibfield  {author} {\bibinfo {author} {\bibfnamefont {M.}~\bibnamefont
  {Ciccolini}}, \bibinfo {author} {\bibfnamefont {A.}~\bibnamefont {Denner}},\
  and\ \bibinfo {author} {\bibfnamefont {S.}~\bibnamefont {Dittmaier}},\
  }\bibfield  {title} {\bibinfo {title} {{Electroweak and QCD corrections to
  Higgs production via vector-boson fusion at the LHC}},\ }\href
  {https://doi.org/10.1103/PhysRevD.77.013002} {\bibfield  {journal} {\bibinfo
  {journal} {Phys. Rev. D}\ }\textbf {\bibinfo {volume} {77}},\ \bibinfo
  {pages} {013002} (\bibinfo {year} {2008})},\ \Eprint
  {https://arxiv.org/abs/0710.4749} {arXiv:0710.4749 [hep-ph]} \BibitemShut
  {NoStop}%
\bibitem [{\citenamefont {Dabelstein}\ and\ \citenamefont
  {Hollik}(1992)}]{Dabelstein:1991ky}%
  \BibitemOpen
  \bibfield  {author} {\bibinfo {author} {\bibfnamefont {A.}~\bibnamefont
  {Dabelstein}}\ and\ \bibinfo {author} {\bibfnamefont {W.}~\bibnamefont
  {Hollik}},\ }\bibfield  {title} {\bibinfo {title} {{Electroweak corrections
  to the fermionic decay width of the standard Higgs boson}},\ }\href
  {https://doi.org/10.1007/BF01625912} {\bibfield  {journal} {\bibinfo
  {journal} {Z. Phys. C}\ }\textbf {\bibinfo {volume} {53}},\ \bibinfo {pages}
  {507} (\bibinfo {year} {1992})}\BibitemShut {NoStop}%
\bibitem [{\citenamefont {Bardin}\ \emph {et~al.}(1991)\citenamefont {Bardin},
  \citenamefont {Vilensky},\ and\ \citenamefont {Khristova}}]{Bardin:1990zj}%
  \BibitemOpen
  \bibfield  {author} {\bibinfo {author} {\bibfnamefont {D.~Y.}\ \bibnamefont
  {Bardin}}, \bibinfo {author} {\bibfnamefont {B.~M.}\ \bibnamefont
  {Vilensky}},\ and\ \bibinfo {author} {\bibfnamefont {P.~K.}\ \bibnamefont
  {Khristova}},\ }\bibfield  {title} {\bibinfo {title} {{Calculation of the
  Higgs boson decay width into fermion pairs}},\ }\href@noop {} {\bibfield
  {journal} {\bibinfo  {journal} {Sov. J. Nucl. Phys.}\ }\textbf {\bibinfo
  {volume} {53}},\ \bibinfo {pages} {152} (\bibinfo {year} {1991})}\BibitemShut
  {NoStop}%
\bibitem [{\citenamefont {Buckley}\ \emph {et~al.}(2015)\citenamefont
  {Buckley}, \citenamefont {Ferrando}, \citenamefont {Lloyd}, \citenamefont
  {Nordstr\"om}, \citenamefont {Page}, \citenamefont {R\"ufenacht},
  \citenamefont {Sch\"onherr},\ and\ \citenamefont {Watt}}]{Buckley:2014ana}%
  \BibitemOpen
  \bibfield  {author} {\bibinfo {author} {\bibfnamefont {A.}~\bibnamefont
  {Buckley}}, \bibinfo {author} {\bibfnamefont {J.}~\bibnamefont {Ferrando}},
  \bibinfo {author} {\bibfnamefont {S.}~\bibnamefont {Lloyd}}, \bibinfo
  {author} {\bibfnamefont {K.}~\bibnamefont {Nordstr\"om}}, \bibinfo {author}
  {\bibfnamefont {B.}~\bibnamefont {Page}}, \bibinfo {author} {\bibfnamefont
  {M.}~\bibnamefont {R\"ufenacht}}, \bibinfo {author} {\bibfnamefont
  {M.}~\bibnamefont {Sch\"onherr}},\ and\ \bibinfo {author} {\bibfnamefont
  {G.}~\bibnamefont {Watt}},\ }\bibfield  {title} {\bibinfo {title} {{LHAPDF6:
  parton density access in the LHC precision era}},\ }\href
  {https://doi.org/10.1140/epjc/s10052-015-3318-8} {\bibfield  {journal}
  {\bibinfo  {journal} {Eur. Phys. J. C}\ }\textbf {\bibinfo {volume} {75}},\
  \bibinfo {pages} {132} (\bibinfo {year} {2015})},\ \Eprint
  {https://arxiv.org/abs/1412.7420} {arXiv:1412.7420 [hep-ph]} \BibitemShut
  {NoStop}%
\bibitem [{\citenamefont {Ball}\ \emph {et~al.}(2017)\citenamefont {Ball} \emph
  {et~al.}}]{NNPDF:2017mvq}%
  \BibitemOpen
  \bibfield  {author} {\bibinfo {author} {\bibfnamefont {R.~D.}\ \bibnamefont
  {Ball}} \emph {et~al.} (\bibinfo {collaboration} {NNPDF}),\ }\bibfield
  {title} {\bibinfo {title} {{Parton distributions from high-precision collider
  data}},\ }\href {https://doi.org/10.1140/epjc/s10052-017-5199-5} {\bibfield
  {journal} {\bibinfo  {journal} {Eur. Phys. J. C}\ }\textbf {\bibinfo {volume}
  {77}},\ \bibinfo {pages} {663} (\bibinfo {year} {2017})},\ \Eprint
  {https://arxiv.org/abs/1706.00428} {arXiv:1706.00428 [hep-ph]} \BibitemShut
  {NoStop}%
\bibitem [{\citenamefont {Chetyrkin}\ \emph {et~al.}(2000)\citenamefont
  {Chetyrkin}, \citenamefont {Kuhn},\ and\ \citenamefont
  {Steinhauser}}]{Chetyrkin:2000yt}%
  \BibitemOpen
  \bibfield  {author} {\bibinfo {author} {\bibfnamefont {K.~G.}\ \bibnamefont
  {Chetyrkin}}, \bibinfo {author} {\bibfnamefont {J.~H.}\ \bibnamefont
  {Kuhn}},\ and\ \bibinfo {author} {\bibfnamefont {M.}~\bibnamefont
  {Steinhauser}},\ }\bibfield  {title} {\bibinfo {title} {{RunDec: A
  Mathematica package for running and decoupling of the strong coupling and
  quark masses}},\ }\href {https://doi.org/10.1016/S0010-4655(00)00155-7}
  {\bibfield  {journal} {\bibinfo  {journal} {Comput. Phys. Commun.}\ }\textbf
  {\bibinfo {volume} {133}},\ \bibinfo {pages} {43} (\bibinfo {year} {2000})},\
  \Eprint {https://arxiv.org/abs/hep-ph/0004189} {arXiv:hep-ph/0004189}
  \BibitemShut {NoStop}%
\bibitem [{\citenamefont {Schmidt}\ and\ \citenamefont
  {Steinhauser}(2012)}]{Schmidt:2012az}%
  \BibitemOpen
  \bibfield  {author} {\bibinfo {author} {\bibfnamefont {B.}~\bibnamefont
  {Schmidt}}\ and\ \bibinfo {author} {\bibfnamefont {M.}~\bibnamefont
  {Steinhauser}},\ }\bibfield  {title} {\bibinfo {title} {{CRunDec: a C++
  package for running and decoupling of the strong coupling and quark
  masses}},\ }\href {https://doi.org/10.1016/j.cpc.2012.03.023} {\bibfield
  {journal} {\bibinfo  {journal} {Comput. Phys. Commun.}\ }\textbf {\bibinfo
  {volume} {183}},\ \bibinfo {pages} {1845} (\bibinfo {year} {2012})},\ \Eprint
  {https://arxiv.org/abs/1201.6149} {arXiv:1201.6149 [hep-ph]} \BibitemShut
  {NoStop}%
\bibitem [{\citenamefont {Herren}\ and\ \citenamefont
  {Steinhauser}(2018)}]{Herren:2017osy}%
  \BibitemOpen
  \bibfield  {author} {\bibinfo {author} {\bibfnamefont {F.}~\bibnamefont
  {Herren}}\ and\ \bibinfo {author} {\bibfnamefont {M.}~\bibnamefont
  {Steinhauser}},\ }\bibfield  {title} {\bibinfo {title} {{Version 3 of RunDec
  and CRunDec}},\ }\href {https://doi.org/10.1016/j.cpc.2017.11.014} {\bibfield
   {journal} {\bibinfo  {journal} {Comput. Phys. Commun.}\ }\textbf {\bibinfo
  {volume} {224}},\ \bibinfo {pages} {333} (\bibinfo {year} {2018})},\ \Eprint
  {https://arxiv.org/abs/1703.03751} {arXiv:1703.03751 [hep-ph]} \BibitemShut
  {NoStop}%
\bibitem [{\citenamefont {Mondini}\ \emph {et~al.}(2019)\citenamefont
  {Mondini}, \citenamefont {Schiavi},\ and\ \citenamefont
  {Williams}}]{Mondini:2019gid}%
  \BibitemOpen
  \bibfield  {author} {\bibinfo {author} {\bibfnamefont {R.}~\bibnamefont
  {Mondini}}, \bibinfo {author} {\bibfnamefont {M.}~\bibnamefont {Schiavi}},\
  and\ \bibinfo {author} {\bibfnamefont {C.}~\bibnamefont {Williams}},\
  }\bibfield  {title} {\bibinfo {title} {{N$^{3}$LO predictions for the decay
  of the Higgs boson to bottom quarks}},\ }\href
  {https://doi.org/10.1007/JHEP06(2019)079} {\bibfield  {journal} {\bibinfo
  {journal} {JHEP}\ }\textbf {\bibinfo {volume} {06}},\ \bibinfo {pages}
  {079}},\ \Eprint {https://arxiv.org/abs/1904.08960} {arXiv:1904.08960
  [hep-ph]} \BibitemShut {NoStop}%
\end{thebibliography}%

\end{document}